\documentstyle[12pt]{article}
\begin{document}
\title{{\bf Minimum-error discrimination between mixed quantum states }}
\author{Daowen Qiu\\
\small{Department of Computer Science, Zhongshan University,
Guangzhou 510275,}\\{\small People's Republic of China}\\
{\small E-mail address: issqdw@mail.sysu.edu.cn}}
\date{  }

\maketitle
\begin{center}
\begin{minipage}{130mm}
\begin{center}{\bf Abstract}\end{center}
{\small  We derive a general lower bound on the minimum-error
probability for {\it ambiguous discrimination} between arbitrary
$m$ mixed quantum states with given prior probabilities. When
$m=2$, this bound is precisely the well-known Helstrom limit.
Also, we give a general lower bound on the minimum-error
probability for discriminating quantum operations. Then we further
analyze how this lower bound is attainable for ambiguous
discrimination of mixed quantum states by presenting necessary and
sufficient conditions related to it. Furthermore, with a
restricted condition, we work out a upper bound on the
minimum-error probability for ambiguous discrimination of mixed
quantum states. Therefore, some sufficient conditions are obtained
for the minimum-error probability attaining this bound. Finally,
under the condition of the minimum-error probability attaining
this bound, we compare the minimum-error probability for {\it
ambiguously} discriminating arbitrary $m$ mixed quantum states
with the optimal failure probability for {\it unambiguously}
discriminating the same states.}


\end{minipage}
\end{center}
\vskip 10mm

\section*{I. Introduction}

Motivated by the study of quantum communication and quantum
cryptography, distinguishing quantum states has become a
fundamental subject in quantum information science [1,2]. This
problem may be roughly described in this manner [1,2,3,4,5,6]:
Suppose that a transmitter, Alice, wants to convey classical
information to a receiver, Bob, using a quantum channel, and Alice
represents the message conveyed as a mixed quantum state that,
with given prior probabilities, belongs to a finite set of mixed
quantum states, say $\{\rho_{1},\rho_{2},\ldots,\rho_{m}\}$; then
Bob identifies the state by a measurement.

As it is known [4,5,6], if the supports of mixed states
$\rho_{1},\rho_{2},\ldots,\rho_{m}$ are not mutually orthogonal,
then Bob can not reliably identify which state Alice has sent,
namely, $\rho_{1},\rho_{2},\ldots,\rho_{m}$ can not be faithfully
distinguished. However, it is always possible to discriminate them
in a probabilistic means. In reality, up to now, various
strategies have been proposed for distinguishing quantum states.
Assume that  mixed states $\rho_{1},\rho_{2},\ldots,\rho_{m}$ have
the {\it a priori} probabilities
$\eta_{1},\eta_{2},\ldots,\eta_{m}$, respectively. In general,
there are three fashions to discriminate them. The first approach
is {\it ambiguous discrimination} [4,5,6], in which inconclusive
outcome is not allowed, and thus error may result. A measurement
for discrimination consists of $m$ measurement operators (e.g.,
positive semidefinite operators) that form a resolution of the
identity on the Hilbert space spanned by the all eigenvectors
corresponding to all nonzero eigenvalues of
$\rho_{1},\rho_{2},\ldots,\rho_{m}$. Much work has been devoted to
devising a measurement maximizing the success probability for
detecting the states [7,8,9,10,11]. The first important result is
the pioneering work by Helstrom [4]---a general expression of the
minimum achievable error probability for distinguishing between
two mixed quantum states. For the case of more than two quantum
states, necessary and sufficient conditions have been derived for
an optimum measurement maximizing the success probability of
correct detection [5,6,8]. However, analytical solutions for an
optimum measurement have been obtained only for some special cases
[11,12,13,14,15], and, as pointed out in [8], obtaining a concrete
expression for an optimum measurement in the general case is a
difficult and unsolved problem.

The second approach is the so-called {\it unambiguous
discrimination} [1,2,16-26], first suggested by Ivanovic, Dicks,
and Peres [16,17,18] for the discrimination of two pure states. In
contrast to ambiguous discrimination, unambiguous discrimination
allows an inconclusive result to be returned, but no error occurs.
In other words, for distinguishing between $m$ mixed states, this
basic idea is to devise a measurement that with a probability
returns an inconclusive result, but, if the measurement returns an
answer, then the answer is fully correct. Therefore, such a
measurement consists of $m+1$ measurement operators, in which a
measurement operator returns an inconclusive outcome. Analytical
solutions for the optimal failure probabilities have been given
for distinguishing between two and three pure states
[16,17,18,19,20,21]. Chefles [22] showed that a set
$\{|\psi\rangle\}$ of pure states is amendable to unambiguous
discrimination if, and only if they are linearly independent. The
optimal unambiguous discrimination between linearly independent
symmetric and equiprobable pure states was solved in [23]. By
means of Lagrange multiplier, Sun {\it et al.} [24] presented a
scheme for calculating the optimal probability of unambiguous
discrimination among linearly independent, nonorthogonal pure
states. A semidefinite programming approach to unambiguous
discrimination between pure states has been investigated in detail
by Eldar [25]. Some upper bounds on the success probability for
unambiguous discrimination between pure states have also been
presented (see [26] and references therein).

We recollect unambiguous discrimination between  mixed quantum
states. For distinguishing between two mixed quantum  states,
general upper and lower bounds have been derived for the optimal
failure probability by  Rudolph {\it et al.} [27], and,
furthermore, for distinguishing between $m$ mixed states, Feng
{\it et al.} [28] obtained a general lower bound on the minimum
failure probability. The analytical results for the optimal
unambiguous discrimination between two mixed quantum states have
been derived by  Raynal {\it et al.} [29],  by Herzog and Bergou
[30], and by Zhou {\it et al.} [31]. More references regarding
unambiguous discrimination of mixed quantum states may be referred
to [32]. (It is also worth mentioning that a universal
programmable quantum device has been designed recently for
unambiguous discrimination of pure states [33], and such a device
can be considered for discriminating mixed states.)

The third strategy for discrimination combines the former two
methods [34,35,36]. That is to say, under the condition that a
fixed probability of inconclusive outcome is allowed to occur, one
tries to determine the minimum achievable probability of errors
for ambiguous discrimination. Chefles {\it et al.} [34] and
Fiur\'{a}\v{s}ek {\it et al.} [35] considered the case of
discriminating pure states, and Eldar [36] dealt with this
discrimination of mixed states. Indeed, by allowing for an
inconclusive result occurring, then one can obtain a higher
probability of correct detection for getting a conclusive result,
than the probability of correct detection attainable without
inconclusive results appearing.

In general, the above discrimination schemes are assumed to have
{\it a priori} probabilities for the states to be discriminated.
Notably, a different scheme recently addressed by D'Ariano {\it et
al.} [37] is the minimax quantum state discrimination strategy, in
which the optimal measurement has been derived for mixed state
discrimination without {\it a priori} probabilities.

In this paper, we deal with ambiguous discrimination between any
$m$ mixed quantum states and compare with unambiguous
discrimination. The main contributions include three aspects:
First we derive a general lower bound on the minimum-error
probability for distinguishing between any $m$ mixed quantum
states. When $m=2$, this lower bound is precisely the well-known
Helstrom limit [4]. Therefore, in the case of discriminating two
mixed states, this bound can always be achieved. By means of the
lower bound, we further give a lower bound on the minimum-error
probability for discriminating quantum operations. Then we further
analyze how this lower bound is attainable for ambiguous
discrimination of mixed  states by presenting necessary and
sufficient conditions related to it. Furthermore, with a
restricted condition, we work out a upper bound on the
minimum-error probability for ambiguous discrimination of mixed
states. Therefore, some sufficient conditions are obtained for the
minimum-error probability attaining this bound. Finally, under the
condition of the minimum-error probability attaining this bound,
we compare the minimum-error probability for {\it ambiguously}
discriminating arbitrary $m$ mixed states with the optimal failure
probability for {\it unambiguously} discriminating the same mixed
states.  When $m=2$, this result has been proved by Herzeg and
Bergou [38].

The remainder of the paper is organized as follows. In Section II,
we derive a lower bound on the minimum-error probability for
ambiguous discrimination between arbitrary $m$ mixed  states. With
this bound, we give a lower bound on the minimum-error probability
for discriminating quantum operations. Then, in Section III, we
further analyze the reachability of this lower bound derived in
Section II, and, in Subsection A, we show some necessary and
sufficient conditions related to it. Furthermore, in Subsection B,
with a restricted condition, we work out a upper bound on the
minimum-error probability.  After that, in Section IV, we deal
with the relation between the minimum-error probability for
ambiguous discrimination of mixed states and the optimal failure
probability for unambiguous discrimination of the same mixed
states. Finally, some concluding remarks are made in Section V.

In general, notation used in this paper will be explained whenever
new symbols appear. Here we first give a denotation that will be
useful in what follows: For any two linear operators $T_{1}$ and
$T_{2}$ on the same Hilbert space ${\cal H}$, we use $T_{1}\perp
T_{2}$ to denote that the supports of $T_{1}$ and $T_{2}$ are
orthogonal. The support of a linear operator $T$ is the subspace
spanned by the all eigenvectors  corresponding to all nonzero
eigenvalues of $T$.

\section*{II. A lower bound on the minimum-error discrimination between mixed quantum states}

Assume that a quantum system is described by a mixed quantum
state, say  $\rho$, drawn from a collection
$\{\rho_{1},\rho_{2},\ldots,\rho_{m}\}$ of mixed quantum states on
an $n$-dimensional complex Hilbert space ${\cal H}$, with the {\it
a priori} probabilities $\eta_{1},\eta_{2},\ldots,\eta_{m}$,
respectively, where $m\leq n$. We assume without loss of
generality that the all eigenvectors of $\rho_{i}$, $1\leq i\leq
m$, span ${\cal H}$, otherwise we consider the spanned subspace
instead of ${\cal H}$. A mixed quantum state $\rho$ is a positive
semidefinite operator with trace 1, denoted $\textrm{Tr}(\rho)=1$.
(Note that a positive semidefinite operator must be a Hermitian
operator [39,40].)  To detect $\rho$, we need to design a
measurement consisting of $m$ positive semidefinite operators, say
$\Pi_{i}$, $1\leq i\leq m$, satisfying the resolution
\begin{equation}
\sum_{i=1}^{m}\Pi_{i}=I,
\end{equation}
where $I$ denotes the identity operator on ${\cal H}$.  By means
of the measurement  $\Pi_{i}$, $1\leq i\leq m$, if the system has
been prepared by $\rho$, then $\textrm{Tr}(\rho\Pi_{i})$ is the
probability to deduce the system being state $\rho_{i}$.
Therefore, the average probability $P$ of correct detecting the
system's state is as follows:
\begin{equation}
P=\sum_{i=1}^{m}\eta_{i}\textrm{Tr}(\rho_{i}\Pi_{i})
\end{equation}
and, the average probability $Q$ of erroneous detection is then as
\begin{equation}
Q=1-P=1-\sum_{i=1}^{m}\eta_{i}\textrm{Tr}(\rho_{i}\Pi_{i}).
\end{equation}
A main objective is to design an optimum measurement that
minimizes the probability of erroneous detection. As mentioned
above, for the case of $m=2$, the optimum detection problem has
been completely solved by Helstrom [4], and the minimum achievable
error probability, say $Q_{A}$, has been presented by the Helstrom
limit [4]
\begin{equation}
Q_{A}=\frac{1}{2}(1-\textrm{Tr}|\eta_{2}\rho_{2}-\eta_{1}\rho_{1}|),
\end{equation}
where $|A|=\sqrt{A^{\dag}A}$ for any linear operator $A$, and
$A^{\dag}$ denotes the conjugate transpose of $A$.

However, for $m>2$, the problem is much more complicated, and, as
indicated above, there has not been a general analytical
expression for the minimum-error probability for ambiguously
distinguishing between arbitrary $m$ mixed  states. To this end,
we show a general analytical solution to a lower bound on the
minimum-error probability for ambiguously distinguishing between
arbitrary $m$ mixed quantum states. Then we will analyze this
bound. We first present a lemma that is useful in the paper.

{\it Lemma 1}. Let $A$ and $B$ be two positive semidefinite
operators. Then $\textrm{Tr}|A-B|\leq
\textrm{Tr}(A)+\textrm{Tr}(B)$, and the equality holds if, and
only if $A\bot B$.

{\it Proof}. Suppose that $A$ and $B$ have the following spectral
decompositions:
\begin{equation}
A=\sum_{i=1}^{k_{1}}\lambda_{i}|\lambda_{i}\rangle\langle\lambda_{i}|,
\end{equation}
\begin{equation}
B=\sum_{j=1}^{k_{2}}\mu_{j}|\mu_{j}\rangle\langle\mu_{j}|,
\end{equation}
where all $\lambda_{i}>0$ and all $\mu_{j}>0$.

If $A\bot B$, then $\langle\lambda_{i}|\mu_{j}\rangle=0$ for
$1\leq i\leq k_{1}$ and $1\leq j\leq k_{2}$, and thus $AB=BA={\bf
0}$, where ${\bf 0}$ denotes a zero operator. In this case, we
obtain
\begin{eqnarray}
|A-B|&=&\sqrt{(A-B)^{\dag}(A-B)}\\
&=&\sqrt{A^{2}+B^{2}-(AB+BA)}\\
&=&\sqrt{A^{2}+B^{2}}\\
 &=&A+B,
\end{eqnarray}
where Eq. (10) is due to  $A\bot B$. Consequently,
$\textrm{Tr}|A-B|= \textrm{Tr}(A)+\textrm{Tr}(B)$.

Before the following proof, we recall some properties of trace
distance and fidelity.  Indeed, as we know from [40], for any
mixed states $\rho$ and $\sigma$, the trace distance
$D(\rho,\sigma)$ and fidelity $F(\rho,\sigma)$ satisfy:
\begin{equation}
D(\rho,\sigma)=\max_{\{E_{m}\}}\frac{1}{2}\sum_{m}|\textrm{Tr}(E_{m}\rho)-\textrm{Tr}(E_{m}\sigma)|,
\end{equation}
and
\begin{equation}
F(\rho,\sigma)=\min_{\{E_{m}\}}\sum_{m}\sqrt{\textrm{Tr}(E_{m}\rho)\textrm{Tr}(E_{m}\sigma)},
\end{equation}
where the maximum and the minimum are over all POVMs $\{E_{m}\}$,
$F(\rho,\sigma)=\textrm{Tr}\sqrt{(\rho^{1/2}\sigma\rho^{1/2})}$
and $D(\rho,\sigma)=\frac{1}{2}\textrm{Tr}|\rho-\sigma|$.  In
fact, in the proof for Eqs. (11,12) in [40], the traces of  $\rho$
and $\sigma$ being one is not involved, and it only utilizes the
positive semidefinite property of $\rho$ and  $\sigma$. Therefore,
for any positive semidefinite operators $A$ and $B$, Eqs. (11,12)
hold as well,  whose proof is only a repeated process step by step
according to those of [40]. In other words, for any positive
semidefinite operators $A$ and $B$, we also have
\begin{equation}
D(A,B)=\max_{\{E_{m}\}}\frac{1}{2}\sum_{m}|\textrm{Tr}(E_{m}A)-\textrm{Tr}(E_{m}B)|,
\end{equation}
and
\begin{equation}
F(A,B)=\min_{\{E_{m}\}}\sum_{m}\sqrt{\textrm{Tr}(E_{m}A)\textrm{Tr}(E_{m}B)},
\end{equation}
where the maximum and the minimum are over all POVMs $\{E_{m}\}$,
$F(A,B)=\textrm{Tr}\sqrt{(A^{1/2}BA^{1/2})}$ and
$D(A,B)=\frac{1}{2}\textrm{Tr}|A-B|$.

Therefore, we always have
\begin{eqnarray}
\textrm{Tr}|A-B|&=&\max_{\{E_{m}\}}\sum_{m}|\textrm{Tr}(E_{m}A)-\textrm{Tr}(E_{m}B)|\\
&\leq&\max_{\{E_{m}\}}(\sum_{m}\textrm{Tr}(E_{m}A)+\sum_{m}\textrm{Tr}(E_{m}B))\\
&=&\textrm{Tr}(A)+\textrm{Tr}(B),
\end{eqnarray}
where Eq. (17) is due to $\sum_{m}E_{m}=I$ for any POVM
$\{E_{m}\}$.

If $\textrm{Tr}|A-B|= \textrm{Tr}(A)+\textrm{Tr}(B)$ holds, we
claim $A\bot B$. Indeed, by means of  Eq. (13) there is a POVM,
say $\{\Pi_{m}\}$ such that
\begin{equation}
D(A,B)=\frac{1}{2}\sum_{m}|\textrm{Tr}(\Pi_{m}A)-\textrm{Tr}(\Pi_{m}B)|,
\end{equation}
from which we have
\begin{eqnarray}
\mathrm{Tr}|A-B|&=&\sum_{m}|\textrm{Tr}(\Pi_{m}A)-\textrm{Tr}(\Pi_{m}B)|\\
&\leq&\sum_{m}\textrm{Tr}(\Pi_{m}A)+\sum_{m}\textrm{Tr}(\Pi_{m}B)\\
&=&\textrm{Tr}(A)+\textrm{Tr}(B).
\end{eqnarray}
Since we assume  $\textrm{Tr}|A-B|=
\textrm{Tr}(A)+\textrm{Tr}(B)$, inequality (20) must be an
equality, which implies that, for each $m$,
$\textrm{Tr}(\Pi_{m}A)=0$ or $\textrm{Tr}(\Pi_{m}B)=0$. Thus, for
each $m$, we have $\textrm{Tr}(\Pi_{m}A)\textrm{Tr}(\Pi_{m}B)=0$.
As a result,
\begin{eqnarray}
F(A,B)&=&\min_{\{E_{m}\}}\sum_{m}\sqrt{\textrm{Tr}(E_{m}A)\textrm{Tr}(E_{m}B)}\\
&\leq&\sum_{m}\sqrt{\textrm{Tr}(\Pi_{m}A)\textrm{Tr}(\Pi_{m}B)}\\
&=&0.
\end{eqnarray}
Consequently, $F(A,B)=0$, i.e.,
$\textrm{Tr}\sqrt{(A^{1/2}BA^{1/2})}=0$. Therefore, due to
$A^{1/2}BA^{1/2}$ being a  positive semidefinite operator,
$A^{1/2}BA^{1/2}$ is a zero operator. Then $A\bot B$ must hold.
Otherwise, there is at least a pair $(i_{0},j_{0})$ such that
\begin{equation}
\langle\lambda_{i_{0}}|\mu_{j_{0}}\rangle\not=0.
\end{equation}
Further, by means of Eqs. (5,6), we have
\begin{eqnarray}
\langle\lambda_{i_{0}}|A^{1/2}BA^{1/2}|\lambda_{i_{0}}\rangle&=&\lambda_{i_{0}}\langle\lambda_{i_{0}}|B|\lambda_{i_{0}}\rangle\\
&\geq&\lambda_{i_{0}}\mu_{j_{0}}|\langle\mu_{j_{0}}|\lambda_{i_{0}}\rangle|^{2}\\
&>&0,
\end{eqnarray}
which contradicts $A^{1/2}BA^{1/2}$ being a zero operator.
Therefore, we have shown that $\textrm{Tr}|A-B|=
\textrm{Tr}(A)+\textrm{Tr}(B)$ implies $A\bot B$.  This has
completed the proof. \hfill $\Box$

Now we present the following theorem.

{\it Theorem 1}. For any $m$ mixed quantum states
$\rho_{1},\rho_{2},\ldots,\rho_{m}$, with the {\it a priori}
probabilities $\eta_{1},\eta_{2},\ldots,\eta_{m}$, respectively,
then the minimum-error probability $Q_{A}$ satisfies
\begin{equation}
Q_{A}\geq \frac{1}{2}(1- \frac{1}{m-1}\sum_{1\leq i<j\leq m}
\textrm{Tr}|\eta_{j}\rho_{j}-\eta_{i}\rho_{i}|).
\end{equation}

{\it Proof}. It suffices to show that the maximum probability, say
$P_{A}$,  of correct detection satisfies
\begin{equation}
P_{A}\leq\frac{1}{2}(1+ \frac{1}{m-1}\sum_{1\leq i<j\leq m}
\textrm{Tr}|\eta_{j}\rho_{j}-\eta_{i}\rho_{i}|).
\end{equation}
For convenience, we first give two symbols: ${\cal
M}=\{\{\Pi_{i}\}_{i=1}^{m}: \sum_{i=1}^{m}\Pi_{i}=I\}$ where
$\Pi_{i}$ are positive semidefinite operators; and we denote
$\Lambda_{ij}=\eta_{j}\rho_{j}-\eta_{i}\rho_{i}$ in this paper.

According to Eqs. (1,2), we know
\begin{equation}
P_{A}=\max_{\{\Pi_{i}\}_{i=1}^{m}}
\sum_{i=1}^{m}\textrm{Tr}(\eta_{i}\rho_{i}\Pi_{i}),
\end{equation}
where the maximization is performed over all POVMs
$\{\Pi_{i}\}_{i=1}^{m}\in {\cal M}$.  By the way, from the
theoretical point of view [6,7,8,9,10], the ``$\max$" does exist
in Eq. (31), so, we can use ``$\max$" instead of ``$\sup$". Of
course, this representation is independent of our proof and
result.

Note that
\begin{equation}
(m-1)\sum_{i=1}^{m}\textrm{Tr}(\eta_{i}\rho_{i}\Pi_{i})+
\sum_{\stackrel{1\leq i<j\leq m}{k\neq i,j}} \left( \eta_i
\textrm{Tr}(\rho_i\Pi_k) \right)=\sum_{1\leq i<j\leq
m}[\eta_{i}+\textrm{Tr}(\Lambda_{ij}\Pi_{j})],
\end{equation}
where $\sum_{i=1}^{m}\eta_{i}=1$ is used. We know that any Hermitian operator $H$ can be represented as the
form $H=A-B$ where $A$ and $B$ are positive semidefinite operators
and $A\perp B$ (i.e., the supports of $A$ and $B$ are orthogonal).
Indeed, the spectral decomposition of $H$ readily verifies this
fact. Since $\Lambda_{ij}$ is Hermitian, we let
\begin{equation}
\Lambda_{ij}=A_{ij}-B_{ij}
\end{equation}
where $A_{ij}$ and $B_{ij}$ are positive semidefinite operators
with $A_{ij}\perp B_{ij}$. In addition, we represent them with
their spectral decomposition forms
\begin{equation}
A_{ij}=\sum_{k}a_{k}^{(ij)}|\phi_{k}^{(ij)}\rangle\langle\phi_{k}^{(ij)}|,
\end{equation}
\begin{equation}
B_{ij}=\sum_{l}b_{l}^{(ij)}|\varphi_{l}^{(ij)}\rangle\langle\varphi_{l}^{(ij)}|,
\end{equation}
where $|\phi_{k}^{(ij)}\rangle$ and $|\varphi_{l}^{(ij)}\rangle$
are mutually orthogonal for all $k$ and $l$, and $a_{k}^{(ij)}$,
$b_{l}^{(ij)}$ are positive real numbers. With Eqs. (32,33,34,35)
we have
\begin{eqnarray}
&&\nonumber\sum_{i=1}^{m}\textrm{Tr}(\eta_{i}\rho_{i}\Pi_{i})\\
&=&\frac{1}{m-1}\sum_{1\leq
i<j\leq m}[\eta_{i}+\textrm{Tr}(A_{ij}\Pi_{j})-\textrm{Tr}
(B_{ij}\Pi_{j})]-\frac{1}{m-1}\sum_{\stackrel{1\leq i<j\leq
m}{k\neq i,j}} \left( \eta_i \textrm{Tr}(\rho_i\Pi_k)
\right)\\
&\leq&\frac{1}{m-1}\sum_{1\leq i<j\leq m}[\eta_{i}+
\sum_{k}a_{k}^{(ij)}\langle\phi_{k}^{(ij)}|\Pi_{j}|\phi_{k}^{(ij)}\rangle
-
\sum_{l}b_{l}^{(ij)}\langle\varphi_{l}^{(ij)}|\Pi_{j}|\varphi_{l}^{(ij)}\rangle]\\
&\leq&\frac{1}{m-1}\sum_{1\leq i<j\leq m}[\eta_{i}+
\sum_{k}a_{k}^{(ij)}],
\end{eqnarray}
where Ineq. (37) is due to $\sum_{\stackrel{1\leq i<j\leq m}{k\neq
i,j}} \left( \eta_i \textrm{Tr}(\rho_i\Pi_k) \right)\geq 0$.  Next
we show that
\begin{equation}
\frac{1}{2}(1+\frac{1}{m-1}\sum_{1\leq i<j\leq m}\textrm{Tr}
|\Lambda_{ij}|)= \frac{1}{m-1}\sum_{1\leq i<j\leq m}(\eta_{i}+
\sum_{k}a_{k}^{(ij)}).
\end{equation}
By combining $\textrm{Tr}(\Lambda_{ij})=\eta_{j}-\eta_{i}$ with
Eqs. (33,34,35), we have
\begin{equation}
\textrm{Tr}(\Lambda_{ij})=\eta_{j}-\eta_{i}=\sum_{k}a_{k}^{(ij)}-\sum_{l}b_{l}^{(ij)}.
\end{equation}
Since  $A_{ij}\perp B_{ij}$, with Lemma 1 and Eqs. (33,34,35) we
further have
\begin{equation}
\textrm{Tr}|\Lambda_{ij}|=\textrm{Tr}(A_{ij})+\textrm{Tr}(B_{ij})=
\sum_{k}a_{k}^{(ij)}+\sum_{l}b_{l}^{(ij)}.
\end{equation}
Therefore, with Eqs. (40,41) we obtain
\begin{eqnarray}
\nonumber && \frac{1}{2}(1+\frac{1}{m-1}\sum_{1\leq i<j\leq
m}\textrm{Tr}|\Lambda_{ij}|)\\
&=&\frac{1}{2}[1+\frac{1}{m-1}\sum_{1\leq i<j\leq
m}(\sum_{k}a_{k}^{(ij)}+\sum_{l}b_{l}^{(ij)})]\\
 &=&\frac{1}{2}[1+\frac{1}{m-1} \sum_{1\leq i<j\leq
m}(2\sum_{k}a_{k}^{(ij)}+\eta_{i}-\eta_{j})]\\
&=&\frac{1}{2}+\frac{1}{m-1} \sum_{1\leq i<j\leq
m}(\eta_{i}+\sum_{k}a_{k}^{(ij)})-\frac{1}{2(m-1)}\sum_{1\leq i<j\leq m}(\eta_{i}+\eta_{j})\\
&=&\frac{1}{m-1}\sum_{1\leq i<j\leq m}(\eta_{i}+
\sum_{k}a_{k}^{(ij)}),
\end{eqnarray}
where the last equality results from
\begin{equation}
\frac{1}{m-1}\sum_{1\leq i<j\leq m}(\eta_{i}+\eta_{j})=1.
\end{equation}
As a result, Eq. (39) holds, and, in terms of Ineq. (38), the
theorem has been proved. \hfill   $\Box$

{\it Remark 1}. With Lemma 1,
$\textrm{Tr}|\eta_{j}\rho_{j}-\eta_{i}\rho_{i}|\leq
\eta_{j}+\eta_{i}$ and, the equality holds if and only if
$\rho_{j}\perp \rho_{i}$.  In Theorem 1, the upper bound on the
probability of correct detection between $m$ mixed quantum states
satisfies
\begin{eqnarray}
 \nonumber &&\frac{1}{2}(1+ \frac{1}{m-1}\sum_{1\leq i<j\leq m}
\textrm{Tr}|\eta_{j}\rho_{j}-\eta_{i}\rho_{i}|)\\
&\leq& \frac{1}{2}[1+ \frac{1}{m-1}\sum_{1\leq i<j\leq m}
(\eta_{j}+\eta_{i}) ]=1,
\end{eqnarray}
and, by means of Lemma 1, we further see that this bound is
strictly smaller than 1 usually unless
$\rho_{1},\rho_{2},\ldots,\rho_{m}$ are mutually orthogonal. \hfill $\Box$

{\it Remark 2}. When $m=2$,  the lower bound in Theorem 1 is
$\frac{1}{2}(1-\textrm{Tr}|\eta_{2}\rho_{2}-\eta_{1}\rho_{1}|)$,
which accords with the well-known Helstrom limit [4]; and indeed,
in this case, this bound can always be attained by choosing the
optimum {\it positive operator-valued measurement} (POVM):
$\Pi_{2}=\sum_{k}|\phi_{k}^{(12)}\rangle\langle \phi_{k}^{(12)}|$
and $\Pi_{1}=I-\Pi_{2}$.  \hfill $\Box$

{\it Remark 3}. From Theorem 1 it readily follows a lower bound on
the minimum-error probability for discriminating $m$ quantum
operations. With respect to quantum operations, we refer to [40].
The problem of the minimum-error discrimination between two
quantum operations, say ${\cal E}_{1}$ and ${\cal E}_{2}$, with
given prior probabilities $\eta_{1}, \eta_{2}$, respectively, has
been formulated by Sacchi [41]. The  minimum-error probability,
say $Q_{E}$, equals
\begin{equation}
Q_{E}=\frac{1}{2}(1-\max_{\rho} \textrm{Tr}|\eta_{2}{\cal
E}_{2}(\rho)-\eta_{1}{\cal E}_{1}(\rho)|)
\end{equation}
where $\rho$ is in the Hilbert space ${\cal H}$ under
consideration.

Then, for arbitrary $m$ quantum operations ${\cal E}_{1}, {\cal
E}_{2},\ldots,{\cal E}_{m}$ with the {\it a priori} probabilities
$\eta_{1}, \eta_{2},\ldots,\eta_{m}$, respectively, in terms of a
POVM $\{\Pi_{i}:1\leq i\leq m\}$, the probability of erroneous
detection is
\begin{equation}
1-\max_{\rho}\sum_{i=1}^{m}\eta_{i}\textrm{Tr}[{\cal
E}_{i}(\rho)\Pi_{i}],
\end{equation}
where $\rho$ is in the Hilbert space ${\cal H}$ under
consideration. Therefore, by means of Theorem 1, the minimum-error
probability $Q_{E}$ for discriminating ${\cal E}_{1}, {\cal
E}_{2},\ldots,{\cal E}_{m}$ satisfies
\begin{equation}
Q_{E}\geq \min_{\rho}\frac{1}{2}(1-\frac{1}{m-1}\sum_{1\leq
i<j\leq m}\textrm{Tr}|\eta_{j}{\cal E}_{j}(\rho)-\eta_{i}{\cal
E}_{i}(\rho)|).
\end{equation}\hfill $\Box$

\section*{III. Further analysis on the lower bound}

In this section, we analyze how the lower bound derived in Section
II can be approached.  As we know, for any POVM $\{\Pi_{j}: 1\leq
j\leq m\}$, $\sum_{\stackrel{1\leq i<j\leq m}{k\neq i,j}} \left(
\eta_i \textrm{Tr}(\rho_i\Pi_k) \right)\geq 0$. When
$\textrm{Tr}(\rho_i\Pi_k)=0$ for all $1\leq i\leq m-1$ and $k\not=
i$, it equals 0, which shows that Ineq. (37) can become an
equality in this case (for example, a strong condition is
 that $\rho_{1},\rho_{2},\ldots,\rho_{m-1}$ are mutually orthogonal).

In Subsection A, we will show in detail that Ineq. (38) in the
proof of Theorem 1 can become an equality if and only if
$\rho_{1},\rho_{2},\ldots,\rho_{m}$ satisfy a certain condition.
Then, in Subsection B, we deal with Ineq. (37) and determine a
upper bound on the minimum-error probability for discriminating
$m$ quantum states under certain conditions.

\subsection*{A. Necessary and sufficient conditions concerning
inequality (38)}

From  Ineq. (38) in the proof of Theorem 1, we can see that this
upper bound for correct detection between $m$ mixed quantum states
can be achieved if, and only if there exists a POVM
$\{\hat{\Pi}_{j}: 1\leq j\leq m\}$ such that
\begin{equation}
\langle\phi_{k}^{(ij)}|\hat{\Pi}_{j}|\phi_{k}^{(ij)}\rangle =1
\end{equation}
and
\begin{equation}
\langle\varphi_{l}^{(ij)}|\hat{\Pi}_{j}|\varphi_{l}^{(ij)}\rangle=0
\end{equation}
for any $1\leq i<j\leq m$, and all $k,l$. In this subsection, we
can clearly formulate this observation and give detailed proof. We
first give the following lemma that is useful to our proof.

{\it Lemma 2}. Let $H$ be a finite dimension Hilbert space. Let
$S$ be a subspace of $H$, and $S$ is spanned by a finite set of
some unit vectors, say $\{|\psi_{j}\rangle: 1\leq j\leq k\}$.
Suppose that $\Pi$ is a positive semidefinite operator on $H$, and
$\Pi\leq I$ (i.e. $I-\Pi$ is a positive semidefinite operator) but
satisfies
\begin{equation}
\langle \psi_{j}|\Pi|\psi_{j}\rangle=1,
\end{equation}
for all $1\leq j\leq k$. Then,
\begin{equation}
\Pi\geq P_{S}
\end{equation}
where $P_{S}$ is a projection operator onto $S$, and $\Pi\geq
P_{S}$ means that $\Pi- P_{S}$ is a positive semidefinite
operator.

{\it Proof}. Since $\Pi$ is a positive semidefinite operator and
$\Pi\leq I$, $\Pi$ has a spectral decomposition of the following
form:
\begin{equation}
\Pi =\sum_{i=1}^{l}a_{i}|a_{i}\rangle\langle a_{i}|
\end{equation}
where $1\geq a_{i}>0$, $1\leq i\leq l$,  and
$\{|a_{i}\rangle:1\leq i\leq l\}$ are orthonormal vectors. By
means of Eqs. (53,55), for any $1\leq j\leq k$, we have
\begin{eqnarray}
1 &=&\langle \psi_{j}|\Pi|\psi_{j}\rangle\\
&=& \sum_{i=1}^{l}a_{i}|\langle\psi_{j}|a_{i}\rangle|^{2}\\
&\leq&\sum_{i=1}^{l}|\langle\psi_{j}|a_{i}\rangle|^{2}\\
&\leq&\langle\psi_{j}|\psi_{j}\rangle\\
&=& 1.
\end{eqnarray}
Therefore, the above inequalities (58,59) must be two equalities.
Consequently,  we obtain
\begin{equation}
\sum_{i=1}^{l}a_{i}|\langle\psi_{j}|a_{i}\rangle|^{2}
=\sum_{i=1}^{l}|\langle\psi_{j}|a_{i}\rangle|^{2}
\end{equation}
and
\begin{equation}
\sum_{i=1}^{l}|\langle\psi_{j}|a_{i}\rangle|^{2}
=\langle\psi_{j}|\psi_{j}\rangle
\end{equation}
for any $1\leq j\leq k$.

If $a_{i}<1$, then from Eq. (61) it follows that
$|\langle\psi_{j}|a_{i}\rangle|=0$ for $1\leq j\leq k$; and, by
combining this result with Eq. (62) we further have
\begin{equation}
\sum_{i=1;a_{i}=1}^{l}|\langle\psi_{j}|a_{i}\rangle|^{2}
=\langle\psi_{j}|\psi_{j}\rangle
\end{equation}
for any $1\leq j\leq k$.

By Eq. (63) we obtain that $|\psi_{j}\rangle$ can be linearly
represented by the vectors in $\{|a_{i}\rangle: a_{i}=1\}$. Since
$S$ is spanned by $\{|\psi_{j}\rangle: 1\leq j\leq k\}$, any
vectors in $S$ can be linearly represented by
$\{|a_{i}\rangle:a_{i}=1\}$. In other words, $\{|a_{i}\rangle:
a_{i}=1\}$ spans a subspace of $H$, say $S_{\Pi}$, satisfying
$S\subseteq S_{\Pi}$. Therefore,
$\sum_{i,a_{i}=1}|a_{i}\rangle\langle a_{i}|$, denoted by
$P_{S_{\Pi}}$, is a projection operator onto $S_{\Pi}$. Meanwhile,
$\sum_{j, a_{j}<1}a_{j}|a_{j}\rangle\langle a_{j}|$, denoted by
$\Pi_{S_{\Pi}^{\perp}}$, is a positive semidefinite operator,
satisfying $\Pi_{S_{\Pi}^{\perp}}\perp P_{S_{\Pi}}$.

Now we rewrite the spectral decomposition of $\Pi$ by regrouping
its items in the following way:
\begin{eqnarray}
\Pi&=& \sum_{i,a_{i}=1}|a_{i}\rangle\langle a_{i}|+
\sum_{j, a_{j}<1}a_{j}|a_{j}\rangle\langle a_{j}|\\
&=& P_{S_{\Pi}}+\Pi_{S_{\Pi}^{\perp}}.
\end{eqnarray}

Since $S\subseteq S_{\Pi}$, we have $P_{S}\leq P_{S_{\Pi}}$ and
therefore $\Pi\geq P_{S_{\Pi}}\geq P_{S}$. This proof has been
completed. \hfill $\Box$

Before giving Theorem 2, we still bring in a couple of symbols.
For any $m$ mixed quantum states
$\rho_{1},\rho_{2},\ldots,\rho_{m}$, with the {\it a priori}
probabilities $\eta_{1},\eta_{2},\ldots,\eta_{m}$, respectively,
where, as before, we assume that the all eigenvectors
corresponding to all nonzero eigenvalues of
$\rho_{1},\rho_{2},\ldots,\rho_{m}$ span an $n$-dimension Hilbert
space ${\cal H}$ ($m\leq n$). Let $S_{ij}^{(+)}$ denote the
subspace spanned by the all eigenvectors corresponding to all {\it
positive} eigenvalues of the Hermitian operator
$\Lambda_{ij}=\eta_{j}\rho_{j}-\eta_{i}\rho_{i}$, and similarly,
$S_{ij}^{(-)}$ represents the subspace spanned by the all
eigenvectors corresponding to all {\it negative} eigenvalues of
 $\Lambda_{ij}$. We use $S_{k}$ to denote the
subspace spanned by the all eigenvectors corresponding to all {\it
positive} eigenvalues of the $k-1$ Hermitian operators
$\Lambda_{1k},\Lambda_{2k},\ldots,\Lambda_{k-1 k}$, $2\leq k\leq
m$. Therefore, $S_{k}$ is the subspace spanned by
$\bigcup_{i=1}^{k-1}S_{ik}^{(+)}$.

With these symbols we present Theorem 2.

{\it Theorem 2}. For any $m$ mixed quantum states
$\rho_{1},\rho_{2},\ldots,\rho_{m}$, with the {\it a priori}
probabilities $\eta_{1},\eta_{2},\ldots,\eta_{m}$, respectively,
then there exists a POVM  $\{\Pi_{i}:1\leq i\leq m\}$ such that
\begin{equation}
\frac{1}{m-1}\sum_{1\leq i<j\leq
m}[\eta_{i}+\textrm{Tr}(\Lambda_{ij}\Pi_{j})]= \frac{1}{2}(1-
\frac{1}{m-1}\sum_{1\leq i<j\leq m}\textrm{Tr}|\Lambda_{ij}|)
\end{equation}
if and only if the following two conditions hold:

(i) For any $1\leq i_{1}, i_{2}<j\leq m$,
\begin{equation}
P_{i_{1}j}^{(+)} \bot P_{i_{2}j}^{(-)}
\end{equation}
where $P_{ij}^{(+)}$ and $P_{ij}^{(-)}$ represent the projection
operators onto  $S_{ij}^{(+)}$ and $S_{ij}^{(-)}$, respectively.

(ii) For any $2\leq i<j\leq m$,
\begin{equation}
P_{i} \bot P_{j}
\end{equation}
where $P_{k}$ denotes the projection operator onto $S_{k}$.

{\it Proof}. (If). First, in terms of the condition (ii) described
by Eq. (68),  we know that
\begin{equation}
\sum_{j=2}^{m}P_{j}\leq I,
\end{equation}
since $S_{k}$ is a subspace of ${\cal H}$, $k=2,3,\ldots, m$, and
they are pairwise orthogonal.

We still use the symbols in the proof of Theorem 1. Recall that
$\Lambda_{ij}=A_{ij}-B_{ij}$ and $A_{ij}\perp B_{ij}$, where $
A_{ij}=\sum_{k}a_{k}^{(ij)}|\phi_{k}^{(ij)}\rangle\langle\phi_{k}^{(ij)}|,
$ and
$B_{ij}=\sum_{l}b_{l}^{(ij)}|\varphi_{l}^{(ij)}\rangle\langle\varphi_{l}^{(ij)}|.
$

 Then, we have
\begin{equation}
P_{ij}^{(+)}
=\sum_{k}|\phi_{k}^{(ij)}\rangle\langle\phi_{k}^{(ij)}|,
\end{equation}
\begin{equation}
 P_{ij}^{(-)} =\sum_{l}|\varphi_{l}^{(ij)}\rangle\langle\varphi_{l}^{(ij)}|.
\end{equation}

According to the condition (i) described by Eq. (67), for any
$1\leq i_{1}, i_{2}<j$, we know that $|\phi_{k}^{(i_{1}j)}\rangle$
and $|\varphi_{l}^{(i_{2}j)}\rangle$ are  orthogonal for all $k$
and $l$.

We know that $S_{j}$ is the subspace spanned by
$\bigcup_{k}\{|\phi_{k}^{(ij)}\rangle:1\leq i<j\}$, $2\leq j\leq
m$. With Eq. (69) we can take a POVM:  $\hat{\Pi}_{j}=P_{j}$ for
$j=2,3,\ldots,m$, and
$\hat{\Pi}_{1}=I-\sum_{j=2}^{m}\hat{\Pi}_{j}$. Then,  for $1\leq
i_{1}<j\leq m$, we have
$\langle\phi_{k}^{(i_{1}j)}|P_{j}|\phi_{k}^{(i_{1}j)}\rangle=1$.
Meanwhile, according to condition (i) described by Eq. (67), we
have
 $\langle\varphi_{l}^{(i_{2}j)}|P_{j}|\varphi_{l}^{(i_{2}j)}\rangle=0$
for $1\leq   i_{2}<j\leq m$.

Therefore, with this POVM, Ineq. (38) in the proof of Theorem 1
will become an equality; more exactly, we obtain that
\begin{eqnarray}
&&\nonumber\frac{1}{m-1}\sum_{1\leq i<j\leq
m}[\eta_{i}+\textrm{Tr}(\Lambda_{ij}\hat{\Pi}_{j})]\\
&=&\frac{1}{m-1}\sum_{1\leq
i<j\leq m}[\eta_{i}+\textrm{Tr}(A_{ij}\hat{\Pi}_{j})-\textrm{Tr}(B_{ij}\hat{\Pi}_{j})]\\
&=&\frac{1}{m-1}\sum_{1\leq i<j\leq m}[\eta_{i}+
\sum_{k}a_{k}^{(ij)}\langle\phi_{k}^{(ij)}|P_{j}|\phi_{k}^{(ij)}\rangle
-
\sum_{l}b_{l}^{(ij)}\langle\varphi_{l}^{(ij)}|P_{j}|\varphi_{l}^{(ij)}\rangle]\\
&=&\frac{1}{m-1}\sum_{1\leq i<j\leq m}[\eta_{i}+
\sum_{k}a_{k}^{(ij)}]\\
&=&\frac{1}{2}(1+ \frac{1}{m-1}\sum_{1\leq i<j\leq n}
\textrm{Tr}|\eta_{j}\rho_{j}-\eta_{i}\rho_{i}|),
\end{eqnarray}
where the last equality results from Eq. (39). As a consequence,
Eq. (66) holds.

(Only if). If there exists a POVM $\{\Pi_{i}:1\leq i\leq m\}$ such
that Eq. (66) holds, then there exists a POVM $\{\hat{\Pi}_{j}:
1\leq j\leq m\}$ such that Eqs. (51,52) hold, that is, $
\langle\phi_{k}^{(ij)}|\hat{\Pi}_{j}|\phi_{k}^{(ij)}\rangle =1 $
and $
\langle\varphi_{l}^{(ij)}|\hat{\Pi}_{j}|\varphi_{l}^{(ij)}\rangle=0
$ for any $1\leq i<j\leq m$, and all $k,l$.  For $j=2,3,\ldots,m$,
$\bigcup_{k}\{|\phi_{k}^{(ij)}\rangle: 1\leq i< j\}$ spans
$S_{j}$,  so, by using Lemma 2, we have
\begin{equation}
\hat{\Pi}_{j}\geq P_{j},
\end{equation}
and, therefore,
\begin{equation}
\sum_{j=2}^{m}\hat{\Pi}_{j}\geq \sum_{j=2}^{m}P_{j}.
\end{equation}
Since $P_{j}$, $2\leq j\leq m$, are some projection operators, we
can conclude that $P_{i}\perp P_{j}$ for $2\leq i<j\leq m$.
Otherwise, if $P_{i_{0}}\perp P_{j_{0}}$ does {\it not} hold for
some $2\leq i_{0}<j_{0}\leq m$, then there exists state
$|\Phi_{i_{0}}\rangle\in S_{i_{0}}$ such that
\begin{equation}
|\Phi_{i_{0}}\rangle=|\Phi_{j_{0}}\rangle +
|\Phi_{j_{0}}^{\perp}\rangle,
\end{equation}
where $0\not= |\Phi_{j_{0}}\rangle\in S_{j_{0}}$, and
$|\Phi_{j_{0}}^{\perp}\rangle$ is orthogonal to $S_{j_{0}}$. Then,
with Ineq. (77) and Eq. (78) we have
\begin{eqnarray}
\langle\Phi_{i_{0}}|\sum_{j=2}^{m}\hat{\Pi}_{j}|\Phi_{i_{0}}\rangle&\geq&
\langle\Phi_{i_{0}}|\sum_{j=2}^{m}P_{j}|\Phi_{i_{0}}\rangle\\
&=&\sum_{j=2}^{m}\langle\Phi_{i_{0}}|P_{j}|\Phi_{i_{0}}\rangle\\
&\geq&\langle\Phi_{i_{0}}|P_{i_{0}}|\Phi_{i_{0}}\rangle+\langle\Phi_{i_{0}}|P_{j_{0}}|\Phi_{i_{0}}\rangle\\
&=&\langle\Phi_{i_{0}}|\Phi_{i_{0}}\rangle+\langle\Phi_{j_{0}}|\Phi_{j_{0}}\rangle\\
&>&\langle\Phi_{i_{0}}|\Phi_{i_{0}}\rangle,
\end{eqnarray}
which contradicts $\sum_{j=2}^{m}\hat{\Pi}_{j}\leq I$. Therefore,
condition (ii) is proved.

Furthermore, we show that condition (i) holds. By combining
$P_{j}\leq \hat{\Pi}_{j}$ with Eq. (52) (i.e., $
\langle\varphi_{l}^{(ij)}|\hat{\Pi}_{j}|\varphi_{l}^{(ij)}\rangle=0
$ for any $1\leq i<j\leq m$ and all $l$), we obtain that
\begin{equation}
\langle\varphi_{l}^{(ij)}|P_{j}|\varphi_{l}^{(ij)}\rangle=0
\end{equation}
for any $1\leq i<j\leq m$ and all $l$.

Let $P_{j}$ ($2\leq j\leq m$) have the following spectral
decomposition:
\begin{equation}
P_{j}=\sum_{t=1}^{N_{j}}|\Phi_{t}^{(j)}\rangle\langle\Phi_{t}^{(j)}|
\end{equation}
where $\{|\Phi_{t}^{(j)}\rangle: 1\leq t\leq N_{j}\}$ is an
orthonormal base of $S_{j}$. From Eq. (84) it follows that
\begin{equation}
\langle\varphi_{l}^{(ij)}|\Phi_{t}^{(j)}\rangle=0
\end{equation}
for any $1\leq i<j\leq m$ and all $l$ and $t$. Since
$|\phi_{k}^{(i^{'}j)}\rangle\in S_{j}$ for any $1\leq i^{'}<j\leq
m$, and $\{|\Phi_{t}^{(j)}\rangle: 1\leq t\leq N_{j}\}$ an
orthonormal base of $S_{j}$, we know that
$|\phi_{k}^{(i^{'}j)}\rangle$ can be linearly represented by
$|\Phi_{t}^{(j)}\rangle$, $1\leq t\leq N_{j}$. Therefore, by Eq.
(86) we obtain
\begin{equation}
\langle\varphi_{l}^{(i_{1}j)}|\phi_{k}^{(i_{2}j)}\rangle=0
\end{equation}
for any $1\leq i_{1},i_{2}< j\leq m$ and all $l$ and $k$. In other
words, condition (i) described by Eq. (67) holds. So far the proof
has been completed. \hfill $\Box$

{\it Remark 4}. When $m=2$, these two conditions described in
Theorem 2 naturally hold, and  $\sum_{\stackrel{1\leq i<j\leq
m}{k\neq i,j}} \left( \eta_i \textrm{Tr}(\rho_i\Pi_k) \right)=0$,
so, in this case, the lower bound can always be achieved, and it
accords with the Helstrom limit [4]. \hfill $\Box$

{\it Remark 5}. By means of Theorem 2, we can precisely work out
the minimum-error probability for ambiguously discriminating
$\rho_{1},\rho_{2},\ldots,\rho_{m}$, with the {\it a priori}
probabilities $\eta_{1},\eta_{2},\ldots,\eta_{m}$, respectively,
if some conditions are restricted. Indeed, we will deal with this
problem in the next subsection. \hfill $\Box$

\subsection*{B. Analysis concerning
inequality (37)}

In Ineq. (37) we leave out the term
$\frac{1}{m-1}\sum_{\stackrel{1\leq i<j\leq m}{k\neq i,j}} \left(
\eta_i \textrm{Tr}(\rho_i\Pi_k) \right)$. In this subsection, in
terms of the conditions described in Theorem 2, we determine the
value on this term. The condition (ii) in Theorem 2 says that
$S_{i}\perp S_{j}$ for $2\leq i<j\leq m$. Here we further assume
that $S_{1}\perp S_{j}$ for $2\leq j\leq m$, as well, where
$S_{1}$ denotes the support of the positive semidefinite operator
$\eta_{1}\rho_{1}$. With this assumption and conditions (i) and
(ii) in Theorem 2, we can calculate
$\frac{1}{m-1}\sum_{\stackrel{1\leq i<j\leq m}{k\neq i,j}} \left(
\eta_i \textrm{Tr}(\rho_i\Pi_k) \right)$, and, also obtain a upper
bound on the  minimum-error probability for ambiguously
discriminating in the following. We describe this result by the
theorem as follows.

{\it Theorem 3.} For any $m$ mixed quantum states
$\rho_{1},\rho_{2},\ldots,\rho_{m}$, with the {\it a priori}
probabilities $\eta_{1},\eta_{2},\ldots,\eta_{m}$, respectively,
if $S_{1}\perp S_{j}$ for $2\leq j\leq m$, where $S_{1}$ denotes
the support of the positive semidefinite operator
$\eta_{1}\rho_{1}$, and the two conditions described in Theorem 2
hold (that is, (i) for any $1\leq i_{1}, i_{2}<j\leq m$, $
P_{i_{1}j}^{(+)} \bot P_{i_{2}j}^{(-)} $ where $P_{ij}^{(+)}$ and
$P_{ij}^{(-)}$ represent the projection operators onto
$S_{ij}^{(+)}$ and $S_{ij}^{(-)}$, respectively; (ii) for any
$2\leq i<j\leq m$, $ P_{i} \bot P_{j} $ where $P_{k}$ denotes the
projection operator onto $S_{k}$), then the minimum-error
probability $Q_{A}$ for ambiguously discriminating
$\rho_{1},\rho_{2},\ldots,\rho_{m}$ satisfies
\begin{equation}
Q_{A}\leq \frac{1}{2}(1- \frac{1}{m-1}\sum_{1\leq i<j\leq m}
\textrm{Tr} |\Lambda_{ij}|)+\frac{1}{2(m-1)}\sum_{2\leq i<j\leq
m}(\eta_{i}+\eta_{1}-\textrm{Tr}|\Lambda_{1i}|).
\end{equation}

{\it Proof}. Firstly, with the assumption that $S_{1}\perp S_{j}$
for $2\leq j\leq m$, we have
\begin{equation}
\textrm{Tr}(\eta_{1}\rho_{1}P_{j})=0
\end{equation}
for $2\leq j\leq m$. Then, for $2\leq i<j\leq m$, due to
$P_{i}\perp P_{j}$ with $1\leq i<j\leq m$, we have
\begin{eqnarray}
0&\leq& \textrm{Tr}(\eta_{i}\rho_{i}P_{j})\\
&=&\textrm{Tr}[(\eta_{i}\rho_{i}-\eta_{1}\rho_{1})P_{j}]+\textrm{Tr}(\eta_{1}\rho_{1}P_{j})\\
&=&\textrm{Tr}[(A_{1i}-B_{1i})P_{j}]\\
&=&\textrm{Tr}(A_{1i}P_{j})-\textrm{Tr}(B_{1i}P_{j})\\
&=&0-\textrm{Tr}(B_{1i}P_{j})\\
&\leq& 0
\end{eqnarray}
which results in
\begin{equation}
\textrm{Tr}(\eta_{i}\rho_{i}P_{j})=0
\end{equation}
for $2\leq i<j\leq m$. In terms of  Eq. (89) and
$\sum_{i=1}^{m}P_{i}=I$ we have
\begin{eqnarray}
\sum_{j=2}^{m}\textrm{Tr}[\eta_{1}\rho_{1}(P_{1}+P_{j})]&=&\sum_{j=2}^{m}\textrm{Tr}(\eta_{1}\rho_{1}P_{1})\\
&=&\sum_{j=2}^{m}\textrm{Tr}[\eta_{1}\rho_{1}(I-\sum_{i= 2}^{m}P_{i})]\\
&=&\sum_{j=2}^{m}\textrm{Tr}(\eta_{1}\rho_{1})\\
&=&(m-1)\eta_{1}.
\end{eqnarray}
With Eqs. (89,96) and
$\textrm{Tr}[(\eta_{i}\rho_{i}-\eta_{1}\rho_{1})P_{i}]=
\sum_{k}a_{k}^{(1i)}$ we have
\begin{equation}
\sum_{2\leq i<j\leq m}
\textrm{Tr}[(\eta_{i}\rho_{i}-\eta_{1}\rho_{1})(P_{i}+P_{j})]=\sum_{2\leq
i<j\leq m}\sum_{k}a_{k}^{(1i)}
\end{equation}
and
\begin{equation}
\sum_{2\leq i<j\leq m}
\textrm{Tr}[\eta_{1}\rho_{1}(P_{i}+P_{j})]=0.
\end{equation}
By the above Eqs. (100,101,102) we obtain that
\begin{eqnarray}
&&\nonumber\sum_{1\leq i<j\leq
m}\textrm{Tr}[\eta_{i}\rho_{i}(P_{i}+P_{j})]\\
&=&\nonumber\sum_{j=2}^{m}\textrm{Tr}[\eta_{1}\rho_{1}(P_{1}+P_{j})]\\
&&+\nonumber\sum_{2\leq i<j\leq m}
\textrm{Tr}[(\eta_{i}\rho_{i}-\eta_{1}\rho_{1})(P_{i}+P_{j})]\\
&&+ \sum_{2\leq
i<j\leq m} \textrm{Tr}[\eta_{1}\rho_{1}(P_{i}+P_{j})]\\
&=&(m-1)\eta_{1}+\sum_{2\leq i<j\leq m}\sum_{k}a_{k}^{(1i)}.
\end{eqnarray}
Due to
\begin{equation}
\textrm{Tr}(\Lambda_{1i})=\sum_{k}a_{k}^{(1i)}-\sum_{l}b_{l}^{(1i)}=\eta_{i}-\eta_{1}
\end{equation}
and
\begin{equation}
\textrm{Tr}|\Lambda_{1i}|=\sum_{k}a_{k}^{(1i)}+\sum_{l}b_{l}^{(1i)}
\end{equation}
we have
\begin{equation}
\sum_{k}a_{k}^{(1i)}=\frac{1}{2}(\eta_{i}-\eta_{1}+\textrm{Tr}|\Lambda_{1i}|).
\end{equation}
Therefore, with Eqs. (104,107) we obtain that
\begin{eqnarray}
&&\nonumber \frac{1}{m-1}\sum_{\stackrel{1\leq i<j\leq m}{k\neq
i,j}} \left(
\eta_i \textrm{Tr}(\rho_i P_k) \right)\\
&=& \frac{1}{m-1}\sum_{1\leq i<j\leq
m}\eta_{i} \textrm{Tr}(I-P_{i}-P_{j})\\
&=& \frac{1}{m-1} \sum_{1\leq i<j\leq
m}\eta_{i}-\frac{1}{m-1}\sum_{1\leq i<j\leq
m}\textrm{Tr}[\eta_{i}\rho_{i}(P_{i}+P_{j})]\\
&=&\frac{1}{m-1} \sum_{1\leq i<j\leq
m}\eta_{i}-\eta_{1}-\frac{1}{2(m-1)}\sum_{2\leq i<j\leq
m}(\eta_{i}-\eta_{1}+\textrm{Tr}|\Lambda_{1i}|)\\
&=& \frac{1}{2(m-1)}\sum_{2\leq i<j\leq
m}(\eta_{i}+\eta_{1}-\textrm{Tr}|\Lambda_{1i}|).
\end{eqnarray}

Therefore, by combining Theorem 2 with Eq. (111), we conclude that
\begin{eqnarray}
\nonumber
&&\sum_{i=1}^{m}\textrm{Tr}(\eta_{i}\rho_{i}P_{i})\\
&=&\frac{1}{m-1}\sum_{1\leq i<j\leq
m}[\eta_{i}+\textrm{Tr}(\Lambda_{ij}P_{j})]-\frac{1}{m-1}\sum_{\stackrel{1\leq
i<j\leq m}{k\neq i,j}} \left( \eta_i \textrm{Tr}(\rho_i P_k)
\right)\\
&=&\frac{1}{2}(1+ \frac{1}{m-1}\sum_{1\leq i<j\leq m}
\textrm{Tr}|\Lambda_{ij}|)-\frac{1}{2(m-1)}\sum_{2\leq i<j\leq
m}(\eta_{i}+\eta_{1}-\textrm{Tr}|\Lambda_{1i}|),
\end{eqnarray}
which is a lower bound on the success probability for ambiguously
discriminating $\{\rho_{i}\}$. Equivalently,
$1-\sum_{i=1}^{m}\textrm{Tr}(\eta_{i}\rho_{i}P_{i})$ is a upper
bound on the minimum-error probability $Q_{A}$ for ambiguously
discriminating $\{\rho_{i}\}$. Therefore, we conclude that Ineq.
(88) holds, and the proof is completed. \hfill  $\Box$

From the proof of Theorem 3 we obtain some sufficient conditions
on the minimum-error probability $Q_{A}$ attaining the lower bound
$\frac{1}{2}(1- \frac{1}{m-1}\sum_{1\leq i<j\leq m}
\textrm{Tr}|\Lambda_{ij}|)$ for ambiguously discriminating
$\{\rho_{i}\}$, which is represented by the following corollary.

 {\it Corollary 1.} For any $m$ mixed quantum states
$\rho_{1},\rho_{2},\ldots,\rho_{m}$, with the {\it a priori}
probabilities $\eta_{1},\eta_{2},\ldots,\eta_{m}$, respectively,
if:
\begin{enumerate}
 \item  $\eta_{i}+\eta_{1}=\textrm{Tr}$$|\Lambda_{1i}|$ for $2\leq i\leq m-1$,

\item for any $1\leq i_{1}, i_{2}<j\leq m$, $ P_{i_{1}j}^{(+)}
\bot P_{i_{2}j}^{(-)} $,

\item for any $2\leq i<j\leq m$, $ P_{i} \bot P_{j} $,

\item for $2\leq j\leq m$, $S_{1}\perp S_{j}$, where $S_{1}$
denotes the support of the positive semidefinite operator
$\eta_{1}\rho_{1}$,
\end{enumerate}
then the minimum-error probability $Q_{A}$ satisfies
\begin{equation}
Q_{A}= \frac{1}{2}(1- \frac{1}{m-1}\sum_{1\leq i<j\leq m}
\textrm{Tr}|\eta_{j}\rho_{j}-\eta_{i}\rho_{i}|). \end{equation}
\hfill $\Box$

\section*{IV. Comparison between ambiguous and unambiguous discriminations for arbitrary $m$ mixed quantum states }

First, we would like to point out that a comparison of POVMs and
projective measurements in the unambiguous and ambiguous cases has
recently been made in [42].  In this section, we compare the
minimum-error probability of ambiguous discrimination to the
inconclusive probability of unambiguous discrimination for any $m$
mixed states under a certain condition.

 For any given $m$ mixed quantum states
$\rho_{1},\rho_{2},\ldots,\rho_{m}$ with the
{\it a priori} probabilities \\
$\eta_{1},\eta_{2},\ldots,\eta_{m}$, respectively, in this
section, we compare the minimum-error probability $Q_{A}$ with the
optimal failure probability, say $Q_{U}$, for unambiguously
discriminating them, under the condition that $Q_{A}$ attains the
lower bound $ \frac{1}{2}(1- \frac{1}{m-1}\sum_{1\leq i<j\leq m}
\textrm{Tr}|\eta_{j}\rho_{j}-\eta_{i}\rho_{i}|)$.  We present the
main result as follows.

{\it Theorem 4}. For any $m$ mixed quantum states
$\rho_{1},\rho_{2},\ldots,\rho_{m}$, with the {\it a priori}
probabilities $\eta_{1},\eta_{2},\ldots,\eta_{m}$, respectively,
if the minimum-error probability  $Q_{A}$ equals
\begin{equation}\frac{1}{2}(1- \frac{1}{m-1}\sum_{1\leq i<j\leq m}
\textrm{Tr}|\eta_{j}\rho_{j}-\eta_{i}\rho_{i}|), \end{equation}
then
\begin{equation}
Q_{U}\geq 2 Q_{A}
\end{equation}
where $Q_{U}$ denotes the optimal failure probability for
unambiguously discriminating $\rho_{1},\rho_{2},\ldots,\rho_{m}$.

{\it Proof}.  Rudolph {\it et al.} [27] proved that a lower bound
on the failure probability $Q_{U}$ for unambiguously
discriminating $\rho_{1},\rho_{2}$, with given  prior
probabilities $\eta_{1},\eta_{2}$, respectively, is
\begin{equation}
Q_{U}\geq 2\sqrt{\eta_{1}\eta_{2}}F(\rho_{1},\rho_{2}),
\end{equation}
where $F(\rho,\sigma)=(\rho^{1/2}\sigma\rho^{1/2})^{1/2}$. A
generalization to the case of $m$ states has been given by Feng
{\it et al.} [28], i.e.,
\begin{equation}
Q_{U}\geq  \sqrt{\frac{m}{m-1}\sum_{i\not=
j}\eta_{i}\eta_{j}F(\rho_{i},\rho_{j})^{2}}.
\end{equation}
In the light of Cauchy-Schwarz inequality, it is easy to get
\begin{eqnarray}
 \sqrt{\frac{m}{m-1}\sum_{i\not=
j}\eta_{i}\eta_{j}F(\rho_{i},\rho_{j})^{2}}&\geq& \frac{1}{m-1}
\sum_{i\not=
j}\sqrt{\eta_{i}\eta_{j}}F(\rho_{i},\rho_{j})\\
&=&\frac{2}{m-1}\sum_{1\leq i<j\leq
m}\sqrt{\eta_{i}\eta_{j}}F(\rho_{i},\rho_{j}).
\end{eqnarray}
Thus, to show that $Q_{U}\geq 2 Q_{A}$, we only need to prove that
\begin{equation}
\frac{2}{m-1}\sum_{1\leq i<j\leq
m}\sqrt{\eta_{i}\eta_{j}}F(\rho_{i},\rho_{j})\geq 2Q_{A}=
1-\frac{1}{m-1}\sum_{1\leq i<j\leq m}\textrm{Tr}|\Lambda_{ij}|,
\end{equation}
where $\Lambda_{ij}=\eta_{j}\rho_{j}-\eta_{i}\rho_{i}$ as before.
Equivalently, it suffices to show that
\begin{equation}
\frac{1}{m-1}\sum_{1\leq i<j\leq
m}\textrm{Tr}|\Lambda_{ij}|+\frac{2}{m-1}\sum_{1\leq i<j\leq
m}\sqrt{\eta_{i}\eta_{j}}F(\rho_{i},\rho_{j})\geq 1.
\end{equation}
In terms of [40,43,44], by choosing an appropriate orthonormal
base $\{|l^{(ij)}\rangle\}$ as the eigenvectors of positive
semidefinite operator
$\rho_{j}^{-1/2}(\rho_{j}^{1/2}\rho_{i}\rho_{j}^{1/2})\rho_{j}^{-1/2}$,
then
\begin{equation}
F(\rho_{i},\rho_{j})=\sum_{l}\sqrt{\langle
l^{(ij)}|\rho_{i}|l^{(ij)}\rangle}\sqrt{\langle
l^{(ij)}|\rho_{j}|l^{(ij)}\rangle}.
\end{equation}
Denote $e_{l}^{(ij)}=\langle l^{(ij)}|\rho_{i}|l^{(ij)}\rangle$
and $f_{l}^{(ij)}=\langle l^{(ij)}|\rho_{j}|l^{(ij)}\rangle$. Due
to $ \sum_{1\leq i<j\leq m}(\eta_{i}+\eta_{j})=m-1, $ we have
\begin{eqnarray}
m-1-2\sum_{1\leq i<j\leq
m}\sqrt{\eta_{i}\eta_{j}}F(\rho_{i},\rho_{j})&=&\sum_{1\leq
i<j\leq
m}[\eta_{i}+\eta_{j}-2\sqrt{\eta_{i}\eta_{j}}F(\rho_{i},\rho_{j})]\\
&=&\sum_{1\leq i<j\leq
m}\sum_{l}(\sqrt{\eta_{i}}\sqrt{e_{l}^{(ij)}}-
\sqrt{\eta_{j}}\sqrt{f_{l}^{(ij)}}  )^{2},
\end{eqnarray}
where
$\textrm{Tr}(\rho_{i})=\sum_{l}e_{l}^{(ij)}=\sum_{l}f_{l}^{(ij)}=\textrm{Tr}(\rho_{j})=1$
is used.

On the other hand, as before, let $\Lambda_{ij}=A_{ij}-B_{ij}$
where  $A_{ij}$ and  $ B_{ij}$ are positive semidefinite
operators, and $A_{ij}\perp B_{ij}$. Then, by Lemma 1 and
$A_{ij}-B_{ij}=\eta_{j}\rho_{j}-\eta_{i}\rho_{i}$, we have
\begin{eqnarray}
\sum_{1\leq i<j\leq m}\textrm{Tr}|\Lambda_{ij}|&=&\sum_{1\leq
i<j\leq
m}\textrm{Tr}|A_{ij}-B_{ij}|\\
&=&\sum_{1\leq i<j\leq
m}\textrm{Tr}(A_{ij}+B_{ij})\\
&=&\sum_{1\leq i<j\leq m}\sum_{l}( \langle
l^{(ij)}|A_{ij}|l^{(ij)}\rangle+\langle
l^{(ij)}|B_{ij}|l^{(ij)}\rangle)\\
&\geq&\sum_{1\leq i<j\leq m}\sum_{l}| \langle
l^{(ij)}|(A_{ij}-B_{ij})|l^{(ij)}\rangle|\\
&=&\sum_{1\leq i<j\leq
m}\sum_{l}|\eta_{j}f_{l}^{(ij)}-\eta_{i}e_{l}^{(ij)}|\\
&\geq&\sum_{1\leq i<j\leq
m}\sum_{l}(\sqrt{\eta_{j}f_{l}^{(ij)}}-\sqrt{\eta_{i}e_{l}^{(ij)}})^{2}\\
&=& m-1-2\sum_{1\leq i<j\leq
m}\sqrt{\eta_{i}\eta_{j}}F(\rho_{i},\rho_{j})
\end{eqnarray}
where the last equality follows from Eq. (125). Therefore, Ineq.
(122) holds, and the proof has been completed. \hfill $\Box$

 Indeed, we can give a simpler method to show
Theorem 4. We need a fact. As we know from [40], for any mixed
states $\rho$ and $\sigma$,
\begin{equation}
1-F(\rho,\sigma)\leq D(\rho,\sigma)
\end{equation}
where
$F(\rho,\sigma)=\textrm{Tr}\sqrt{(\rho^{1/2}\sigma\rho^{1/2})}$
and $D(\rho,\sigma)=\frac{1}{2}\textrm{Tr}|\rho-\sigma|$.  In
fact, in the proof for Ineq. (133) in [40], the traces of  $\rho$
and $\sigma$ being one is not involved, and it only utilizes the
positive semidefinite property of $\rho$ and  $\sigma$. Therefore,
it follows the following fact, whose proof is only a repeated
process step by step according to those of [40].

{\it Fact 1.} For any two positive semidefinite operators $\rho$
and $\sigma$, we have
\begin{equation}
\frac{\textrm{Tr}(\rho)+\textrm{Tr}(\sigma)}{2}-F(\rho,\sigma)\leq
D(\rho,\sigma),
\end{equation}
where, as above,
$F(\rho,\sigma)=\textrm{Tr}\sqrt{(\rho^{1/2}\sigma\rho^{1/2})}$
and $D(\rho,\sigma)=\frac{1}{2}\textrm{Tr}|\rho-\sigma|$.

{\it Alternative Method for the Proof of Theorem 4:} In the light
of inequalities (118,119), Eq. (120) and Fact 1, we get that
\begin{eqnarray}
 Q_{U}&\geq&\sqrt{\frac{m}{m-1}\sum_{i\not=
j}\eta_{i}\eta_{j}F(\rho_{i},\rho_{j})^{2}}\\
&\geq& \frac{2}{m-1}\sum_{1\leq i<j\leq
m}\sqrt{\eta_{i}\eta_{j}}F(\rho_{i},\rho_{j})\\
&=&\frac{2}{m-1}\sum_{1\leq i<j\leq
m}F(\eta_{i}\rho_{i},\eta_{j}\rho_{j})\\
&\geq&\frac{2}{m-1}\sum_{1\leq i<j\leq
m}(\frac{\eta_{i}+\eta_{j}}{2}-
D(\eta_{i}\rho_{i},\eta_{j}\rho_{j}))\\
&=&1-\frac{1}{m-1}\sum_{1\leq i<j\leq m}\textrm{Tr}|\Lambda_{ij}|\\
&=&2Q_{A},
\end{eqnarray}
where Ineq. (138) is resulted from Fact 1. \hfill $\Box$

\section*{V. Concluding Remarks}

It is a difficult problem for giving an analytical solution for
ambiguously distinguishing between any  $m$ given mixed states,
and only some special cases has been solved [11,12,13,14,15]. In
this paper, we have derived an analytical expression of the lower
bound on the minimum-error probability for ambiguously
distinguishing between arbitrary $m$ mixed  states. When $m=2$,
this bound is precisely the well-known Helstrom limit [4]. Also,
we have provided a lower bound on the minimum-error probability
for discriminating quantum operations. Then we have further
analyzed this lower bound for ambiguous discrimination of mixed
states by presenting necessary and sufficient conditions related
to it. Furthermore, with a restricted condition, we have worked
out a upper bound on the minimum-error probability for ambiguous
discrimination of mixed  states. Therefore, some sufficient
conditions have been presented for the minimum-error probability
attaining this bound. Finally, under the condition that the
minimum-error probability attains this bound, we have compared the
minimum-error probability for {\it ambiguously} discriminating
arbitrary $m$ mixed states with the optimal failure probability
for {\it unambiguously} discriminating the same mixed states.
When $m=2$, this relation has been proved by Herzeg and Bergou
[38].

A further question worthy of consideration is comparison between
unambiguous and ambiguous discriminations without any restricted
conditions. Also, this lower bound we derived may be appropriately
improved, since inequality (21) can be strict for some
$\rho_{1},\rho_{2},\ldots,\rho_{m}$. Based on the paper, another
issue is to further investigate the minimum-error probability for
distinguishing between quantum operations [41]. We would like to
study them in the subsequent work.

\section*{Acknowledgements}
This work is supported by the National Natural Science Foundation
(Nos. 90303024, 60573006), the Research Foundation for the
Doctoral Program of Higher School of Ministry of Education (No.
20050558015), and NCET of China.


\begin{thebibliography}{AB}


\bibitem{ab} A. Chefles,   Contemp.  Phys. {\bf 41,} 401 (2000).
\bibitem{ab}  J.A. Bergou, U. Herzog, and M. Hillery,
 {\it Quantum State Estimation}, Lecture Notes in Physics Vol. 649
(Springer, Berlin, 2004), p. 417; A. Chefles, {\it ibid}. p. 467.

\bibitem{ab} Y.C. Eldar and G.D. Forney, Jr., IEEE Trans. Inform. Theory {\bf 47,} 858
(2001).

\bibitem{ab}C. W. Helstrom, {\it Quantum Detection and Estimation Theory}
(Academic Press, New York, 1976).


\bibitem{ab}A.S. Holevo, J. Multivariate Anal. {\bf 3,} 337
(1973).

\bibitem{ab}H.P. Yuen, R.S. Kennedy, and M. Lax, IEEE Trans. Inform. Theory {\bf IT-21,}
125 (1975).



\bibitem{ab}M. Charbit, C. Bendjaballah, and C. W. Helstrom, IEEE Trans. Inform. Theory {\bf 35,}
1131 (1989).
\bibitem{ab} Y.C. Eldar, A. Megretski, and G.C. Verghess, IEEE Trans. Inform. Theory {\bf 49,}
1007 (2003).
\bibitem{ab}M. Osaki, M. Ban, and O. Hirota, Phys. Rev. A {\bf
54,} 1691 (1996).

\bibitem{ab} M. Ban, K. Kurokawa, R. Momose, and O. Hirota, Int. J. Theor. Phys. {\bf
36,} 1269 (1997).


\bibitem{ab} Y.C. Eldar and G.D. Forney, Jr., e-print arXiv:
quant-ph/0211111.

\bibitem{ab} S.M. Barnett,  Phys. Rev. A {\bf
64,} 030303(R) (2001).


\bibitem{ab} E. Andersson, S.M. Barnett, C.R. Gilson, and K. Hunter,  Phys. Rev. A {\bf
65,} 052308 (2002).


\bibitem{ab} C.-L. Chou and L.Y. Hsu, Phys. Rev. A {\bf
68,} 042305 (2003).


\bibitem{ab} U. Herzog and J.A.  Bergou, Phys. Rev. A {\bf
65,} 050305(R) (2002).

\bibitem{ab} I. D. Ivanovic, Phys. Lett. {\bf A123}, 257 (1987).
\bibitem{ab} D. Dieks, Phys. Lett. {\bf A126}, 303 (1988).
\bibitem{ab} A. Peres, Phys. Lett. {\bf A128}, 19 (1988).
\bibitem{ab} G. Jaeger and A. Shimony, Phys. Lett. {\bf A197}, 83 (1995).
\bibitem{ab}A. Peres and D.R. Terno, J. Phys. A {\bf 31,} 7105
(1998).
\bibitem{ab}L.M. Duan and G.C. Guo, Phys. Rev. Lett.  {\bf 80,} 4999 (1998); C.W. Zhang, C.F. Li, and G.C. Guo, Phys. Lett. A  {\bf 261,} 25 (1999).
\bibitem{ab}A. Chefles, Phys. Lett. A  {\bf 239,} 339  (1998).

\bibitem{ab} A. Chefles, S.M. Barnett, Phys. Lett. A  {\bf 250,}  223 (1998).

\bibitem{ab}Y. Sun, J.A. Bergou, and M. Hillery,  Phys. Rev. A {\bf
66,} 032315 (2002).

\bibitem{ab} Y.C. Eldar,  IEEE Trans. Inform. Theory {\bf 49,}
446 (2003).

\bibitem{ab}  D. Qiu, Phy.
Lett. A {\bf 303,} 140 (2002); D. Qiu, Phy.
Lett. A {\bf 309,} 189 (2003); D. Qiu, J. Phys. A: Math. Gen. {\bf
35,} 6931 (2002).


\bibitem{ab} T. Rudolph, R.W. Spekkens, and P.S. Turner, Phys. Rev. A {\bf
68,} 010301(R) (2003).
\bibitem{ab} Y. Feng, R.Y. Duan, and Z. Ji,  Phys. Rev. A {\bf 72,} 012313
(2005).


\bibitem{ab}P. Raynal, N. L\"{u}tkenhaus, and S.J. van Enk, Phys. Rev. A {\bf
68,} 022308 (2003).


\bibitem{ab} U. Herzog and J.A.  Bergou, Phys. Rev. A {\bf 71,} 050301(R)
(2005).

\bibitem{ab}X.-F. Zhou, Y.-S. Zhang, and G.C. Guo,  Phys. Rev. A {\bf 75,} 052314
(2007).


\bibitem{ab}U. Herzog, Phys. Rev. A {\bf
75,} 052309 (2007).

\bibitem{ab}  J.A. Bergou and M. Hillery, Phys. Rev. Lett. {\bf 94,} 160501 (2005).

\bibitem{ab} A. Chefles and S.M. Barnett, J. Mod. Opt. {\bf 45,} 1295 (1998).

\bibitem{ab} J. Fiur\'{a}\v{s}ek, M. Je\v{z}ek, Phys. Rev. A {\bf
67,} 012321 (2003).

\bibitem{ab} Y.C. Eldar, Phys. Rev. A {\bf
67,} 042309 (2003).


\bibitem{ab}G.M. D'Ariano, M.F. Sacchi, and J. Kahn,  Phys. Rev. A {\bf 72,} 032310 (2005).


\bibitem{ab}U. Herzog and J.A.  Bergou, Phys. Rev. A {\bf 70,} 022302 (2004).

\bibitem{Horn} R.A. Horn, C.R. Johnson, {\it Matrix Analysis}
(Cambridge University Press, Cambridge, 1986).

\bibitem{ab}M.A. Nielsen and I.L. Chuang, {\it Quantum Computation and Quantum Information}
  (Cambridge University Press, Cambridge, 2000).
\bibitem {ab} M.F. Sacchi,  Phys. Rev. A {\bf 71,} 062340 (2005).

\bibitem{ab} M.A.P. Touzel, R.B.A. Adamson,  and A.M. Steinberg, e-print arXiv:
0708.1540v2.


\bibitem{ab} C.A. Fuchs, PhD thesis, Univ. of New Mexico (1995), e-print arXiv: quant-ph/9601020.


\bibitem{ab}H. Barnum, C.M. Caves, C.A. Fuchs, R. Jozsa, and B. Schumacher,
Phys. Rev.  Lett. {\bf 76,} 2818 (1996).












\end{thebibliography}
\end{document}